\documentclass[journal]{IEEEtran}
\usepackage{blindtext}
\usepackage{graphicx}
\usepackage[utf8]{inputenc}
\usepackage{graphics} 
\usepackage{epsfig} 
\usepackage{mathptmx} 
\usepackage{amsmath} 
\usepackage{amssymb}  
\usepackage{cite}
\usepackage{multicol}
\usepackage{mathtools, cuted}
\usepackage[cmintegrals]{newtxmath}
\usepackage{epstopdf}
\usepackage{graphics} 
\usepackage{epsfig} 

\usepackage{mathptmx} 
\usepackage{amsmath} 
\usepackage{amssymb}  
\usepackage{cite}
\usepackage{multicol}
\usepackage{mathtools, cuted}
\usepackage[cmintegrals]{newtxmath}
\usepackage{mathtools}

\usepackage{mathptmx}
\usepackage{helvet}
\usepackage{courier}
\usepackage{subfigure}
\usepackage{type1cm}
\usepackage{titlesec}
\usepackage{hyperref}
\usepackage{siunitx}
\usepackage{stackengine}
\usepackage{epsfig} 
\usepackage{mathptmx} 
\usepackage{amsmath} 
\usepackage{amssymb}  
\usepackage{cite}
\usepackage{multicol}
\usepackage{mathtools, cuted}
\usepackage[cmintegrals]{newtxmath}
\usepackage{makeidx}         
\usepackage{algorithm}
\usepackage{algpseudocode}

\usepackage{multicol}        
\usepackage[bottom]{footmisc}
\usepackage{caption}
\usepackage{nomencl}
\usepackage[usenames, dvipsnames]{color}

\usepackage{multicol}        
\usepackage[bottom]{footmisc}
\usepackage{caption}
\usepackage{titlesec}
\usepackage{nomencl}
\usepackage[usenames, dvipsnames]{color}
\usepackage{multirow}

\usepackage{amsthm}
 
\theoremstyle{definition}
\newtheorem{definition}{Definition}

\theoremstyle{theorem}
\newtheorem{theorem}{Theorem}

\theoremstyle{remark}
\newtheorem{remark}{Remark}
\theoremstyle{proposition}

\theoremstyle{corollary}

\theoremstyle{proof}

\newtheorem{assumption}{Assumption}
\theoremstyle{assumption}

\theoremstyle{lemma}

\ifCLASSINFOpdf
\else
\fi
\begin{document}
%
\title{Multi-Agent Coordination Fluid Flow Modeling and Experimental Evaluation}
%
%
%

\author{ Harshvardhan Uppaluru, Mohammad Ghufran, and Hossein Rastgoftar 
\thanks{{\color{black}Authors are with the Department
of Aerospace and Mechanical Engineering, University of Arizona, Tucson,
AZ, 85721 USA e-mail: \{huppaluru, ghufran1942, hrastgoftar\}@arizona.edu.}}
}
%
%

\markboth{
}%
{Shell \MakeLowercase{\textit{et al.}}: Bare Demo of IEEEtran.cls for IEEE Journals}
%



\maketitle

\begin{abstract}
Reliability is a critical aspect of multi-agent system coordination as it guarantees the system's accurate and consistent functionality. If one agent in the system fails or behaves unexpectedly, it can negatively impact the performance and effectiveness of the entire system. Therefore, it is important to design and implement multi-agent systems with a high level of reliability to ensure that they can operate safely and move smoothly in the presence of unforeseen agent failure or lack of communication with some agent teams moving in a shared motion space. This paper presents a novel  navigation model that, in an ideal fluid-flow, divides agents into cooperative (non-singular) and non-cooperative (singular) agents, with cooperative agents sliding along streamlines safely enclosing non-cooperative agents in a shared motion space. A series of flight experiments utilizing crazyflie quadcopters will experimentally validate the suggested model.
\end{abstract}


\vspace{-0.5cm}
\section{Introduction}
\label{sec:introduction}
Robotics research has long drawn inspiration from nature. Researchers have examined how animals move, communicate, and interact with their environments in order to build robots capable of performing similar tasks. 
Biomimicry, the idea of designing and building technology inspired by nature, has resulted in the development of efficient robots that can adapt in response to their environment. 
Our inspiration stems from the flow of a fluid around a rock offering a glimpse into how robots can navigate around obstacles in their environments. 
For example, the principles of fluid dynamics can be applied to the design of a robot's movement, allowing it to move smoothly and efficiently around obstacles.
\vspace{-0.5cm}
\subsection{Related Work}
Multi-Agent Systems (MAS) have been deployed in a plethora of robotics applications such as search and rescue missions \cite{drew2021multi}, forest robotics \cite{oliveira2021advances}, and surveillance \cite{acevedo2014decentralized}, 
due to their significant advantages 
when compared to a single agent. Such MAS must be equipped with robust algorithms that can safely navigate around both static and dynamic obstacles in order for all agents to successfully complete the cooperative job. Various collision-free path planning works have been previously published such as collision cone \cite{sunkara2019collision}, 
navigation functions \cite{tanner2012multiagent}, velocity obstacle concept \cite{van2008reciprocal, van2010optimal}, flocking \cite{inacio2018united} and sampling based methods \cite{ichter2018learning}. Flow-based control strategies for marine robots in gyre-like flows have been previously studied \cite{knizhnik2022flow, knizhnik2023flow}. 
Artificial Potential Fields (APF)\cite{khatib1986real, kim2011path, chen2009evolutionary} is a simple and mathematically elegant technique originally proposed for manipulators and mobile robots in an operational space. Combining a positive potential around goal location and a negative potential around obstacles, this method guides the robot toward its goal by following the gradients of potential field while steering away from obstacles. A well-known issue of such an approach is getting trapped in local minima and in a real-world dynamic environment it has been shown that APF is inefficient \cite{montiel2015path}. 

Control Barrier Functions (CBFs) have emerged as a potential mathematical tool for safety assurances \cite{jankovic2018robust}.
Using system dynamics, CBFs can be used to define a admissible region in the robot's workspace, and the robot's control inputs are then calculated to ensure that the robot's state remains within this region at all times. CBFs for a safe behavior in multi-agent robotics was studied previously \cite{borrmann2015control} and a decentralized supervisory controller based on CBFs has been presented \cite{chen2020guaranteed}. A combination of CBFs with Control Lyapunov Functions (CLFs) via quadratic programming was studied for cruise control applications \cite{ames2014control}. 
We have recently developed an advanced physics-based automation system for the safe and efficient coordination of large-scale multi-agent systems, even in the face of disturbances and unexpected failures \cite{rastgoftar2019physics, rastgoftar2020fault, UPPALURU2022107960, romano2022quadrotor}. This innovative approach is composed of two operation modes: Homogeneous Deformation Mode (HDM) and Failure Resilient Mode (FRM). By applying the principles of continuum mechanics, we have successfully formalized the transitions between these two modes, enabling a robust response to varying operating conditions.

\subsection{Contributions}
This work presents a novel approach to ensuring the safe and resilient coordination of multiple agent teams moving collectively in a shared motion environment. Drawing inspiration from ideal fluid-flow models, each team treats its agents as cooperative particles within an ideal fluid-flow field while considering other teams' agents as singular points in the field. To ensure inter-agent collision avoidance and safely wrap the non-cooperative agents, the cooperative agents slide along the streamlines of an ideal fluid-flow field. The proposed approach will be experimentally validated using Crazyflie quadcopters in an indoor flight space. Compared to existing literature and the authors' previous work, this paper offers the following key contributions:

\textbf{Contribution 1:} The work extends the experimental evaluation of  navigation presented in~\cite{romano2022quadrotor}, which investigated a single failed quadcopter, by modeling and experimentally validating  navigation in multi-agent systems in the presence of multiple non-concurrent failures and obstacles with arbitrary sizes and geometries.

 \textbf{Contribution 2:} The proposed  navigation approach establishes a novel paradigm for collision avoidance, wherein: (i) each obstacle is treated as a rigid body whose boundary is determined by a streamline enclosing it; and (ii) collision avoidance is ensured by defining agents desired trajectories along the  streamlines that safely wrap obstacles.
 
\textbf{Contribution 3:}  The work models and experimentally validates navigation for multiple agent teams simultaneously coordinating within a shared motion space. In particular, this work presents algorithmic approaches for navigation in the presence of stationary and dynamic obstacles, encompassing various situations such as Stationary Non-Concurrent Failures (SNCF), Time-Varying  Non-Cooperative (TVNC), Time-Varying Cooperative (TVC), and Stationary Obstacle-Laden Environments (SOLE) containing many obstacles with arbitrary sizes and geometries that are randomly distributed. 

\textbf{Contribution 4:} The proposed SOLE fluid-flow navigation approach applies the existing mesh generation techniques \cite{hoffmann1993computational}, mainly used in computational fluid dynamics, to convert a highly-constrained motion space, populated with a random number of obstacles of arbitrary size and geometry into an obstacle-free planning space, and ensure collision avoidance by planning the agent coordination in the planning space. To the best of authors' knowledge, this is the first work that leverages computational fluid dynamics (CFD) mesh generation principles to ensure collision-free multi-agent coordination within a highly-constrained motion environment.


\vspace{-0.3cm}
\subsection{Organization}
The {\color{black}remaining sections} of the paper {\color{black}are} organized as follows: {\color{black}A} detailed description of our proposed methods {\color{black}is} presented in Section \ref{sec:methodology}. The proposed model will be used in Section \ref{Operation Modes} to present five different  operation modes under different communication and constraint protocols.  Section \ref{sec:ExperimentsAndDiscussion} {\color{black}outlines} the experimental setup and {\color{black}presents} the results of the experiments. We finally conclude the paper in Section \ref{sec:conclusion} with thoughts about future {\color{black}directions}.
\section{Methodology}
\label{sec:methodology}
We consider a MAS {\color{black}represented} by {\color{black}the} set $\mathcal{V}=\left\{1,\cdots,N\right\}$, {\color{black}which is subsequently clustered} into $m$ {\color{black}distinct} groups{\color{black}. These groups} are {\color{black}identified by the} set $\mathcal{M}=\left\{1,\cdots,m\right\}$. Let $\mathcal{V}_l$ be a set defining  agents of cluster $l\in \mathcal{M}${\color{black}; consequently, the set} $\bar{\mathcal{V}}_l=\mathcal{V}\setminus \mathcal{V}_l$ defines agents that do not belong to cluster $l\in \mathcal{M}$
{\color{black} Although,} $\mathcal{V}$ {\color{black}remains} time-invariant, the number of agents {\color{black} in} $\mathcal{V}_l$ can {\color{black}vary} with time{\color{black}, suggesting }that $\mathcal{V}_l$ may lose or absorb agents at any {\color{black}given} time $t$. 

To safely plan coordination of $\mathcal{V}_l$'s agents, in the presence of {\color{black}agents belonging to }$\bar{\mathcal{V}}_l$, we consider $\mathcal{V}_l$'s agents {\color{black}as} finite number of particles of an ideal fluid-flow  field while $\bar{\mathcal{V}}_l$'s agents are either considered as ``singularity points'' or ``rigid bodies'' that are safely wrapped by the  streamlines. {\color{black} For the  ideal fluid-flow  field, used for planning of coordination of $\mathcal{V}_l$'s agents, we define potential filed $\phi_l\left(x,y,\theta_l,t\right)$ and stream field $\psi_l\left(x,y,\theta_l(t),t\right)$, where $x$ and $y$ are position components, $t$ denotes time, and $\theta_l(t)$ is the bulk motion direction of cluster $l\in \mathcal{M}$.} Note that both potential and stream function{\color{black}s} satisfy the Laplace equation:
\vspace{-0.3cm}
\begin{subequations}
    \begin{equation}
        \label{phiiiii}
        {\phi_l}_{_{xx}}+ {\phi_l}_{_{yy}}=0,\qquad \forall l\in \mathcal{M}
    \end{equation}
    \begin{equation}
        \label{psiiiii}
        {\color{black}{\psi_l}_{_{xx}}+ {\psi_l}_{_{yy}}=0},\qquad \forall l\in \mathcal{M}
    \end{equation}
\end{subequations}

 For the sake of simplicity, we use $\phi_{il}(t)$ and $\psi_{il}(t)$ to denote $\phi_{il}(t)=\phi_l(x_i,y_i,\theta_l,t)$ and $\psi_{il}(t)=\psi_l(x_i,y_i,\theta_l,t)$ to specify the corresponding  potential and stream coordinate{\color{black}s} of agent $i\in \mathcal{V}_l$ positioned at $(x_i,y_i)$ at time $t$.

\textbf{Path Planning Strategy:} Every agent  $i\in \mathcal{V}_l$ can avoid  inter-agent collision and hitting $\bar{\mathcal{V}}_l$'s agents when it slides along level curve $\psi_l\left(x_i,y_i,\theta_l(t),t\right)=\bar{\psi}_{i,l}(t)$ constant \cite{UPPALURU2022107960}. Therefore, the tangent vector to the desired sliding path of  agent $i\in \mathcal{V}_l$ is obtained by
\vspace{-0.25cm}
\begin{equation}\label{streamline}
    \hat{\mathbf{T}}_i(x,y,t)=\begin{bmatrix}
        \dfrac{\partial \psi_l\left(x_i,y_i,\theta_l,t\right)}{\partial y}&- \dfrac{\partial \psi_l\left(x_i,y_i,\theta_l,t\right)}{\partial x}
    \end{bmatrix}
    ^T,
\end{equation}
for every $i\in \mathcal{V}_l$ and $l\in \mathcal{M}$ (See Fig. \ref{connectionschematic}). for every $i\in \mathcal{V}_l$ and $l\in \mathcal{M}$ (See Fig. \ref{connectionschematic}). 
\begin{figure}
    \centering
    \includegraphics[width=\linewidth]{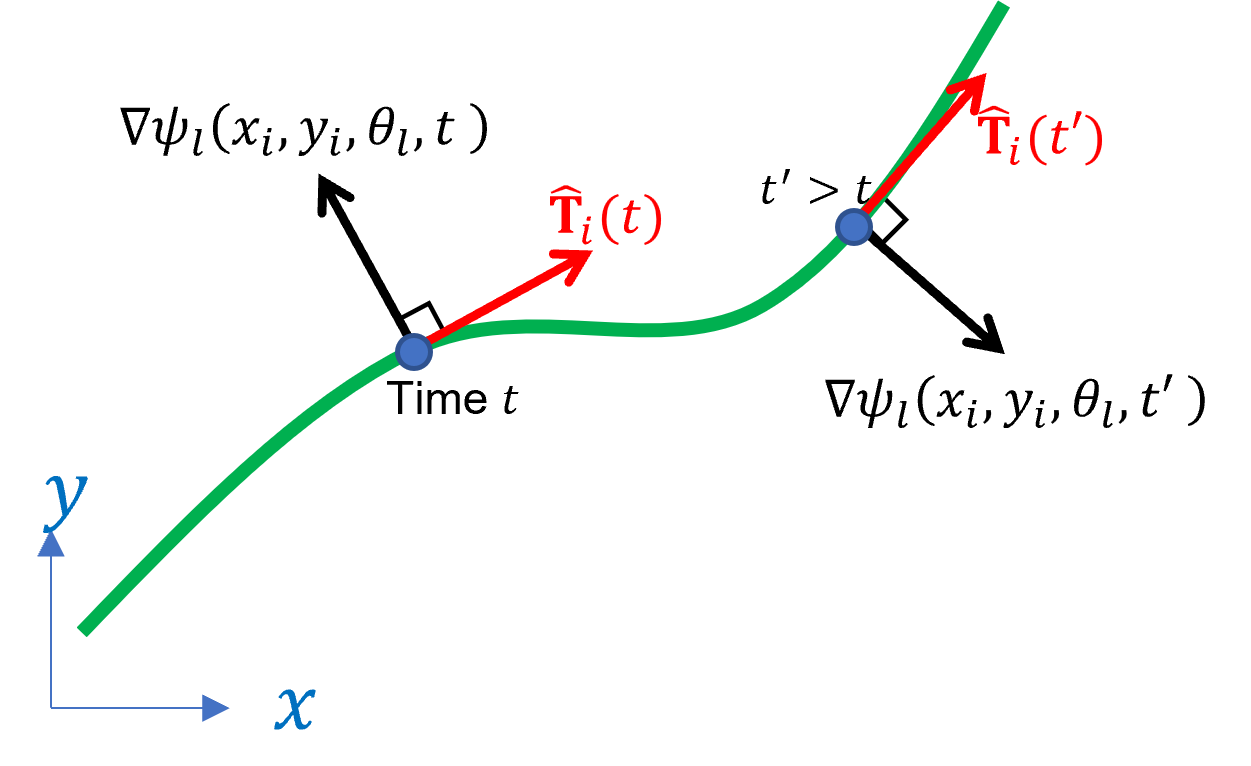}
    \vspace{-0.7cm}
    \caption{Schematic of the desired path of agent $i\in \mathcal{V}_l${\color{black}.}}
    \label{connectionschematic}
\end{figure}
The desired velocity of agent $i\in \mathcal{V}_l$ is given by
\vspace{-0.25cm}
\begin{equation}\label{Eqqqqqq4}
    \mathbf{V}_i=v_l\hat{\mathbf{T}}_i,\qquad \forall i\in \mathcal{V}_l,~l\in \mathcal{M},
\end{equation}
and we use the Algorithm \ref{alg1} to update $\mathbf{z}_i$ at any time $t$.

\begin{theorem}
    \label{theorem1}
    Let $\left(x_{i,0}, y_{i,0}\right)$ denote the position of agent $i \in \mathcal{V}_l$ in the $x-y$ plane and $\theta_l(t_0) = \theta_{l0}$ denote the bulk motion direction of agents in $\mathcal{V}_l$ at initial (reference) time $t_0$ for every $l\in \mathcal{M}$. 
    Define $\phi_{il}(t_0) = \phi_{il,0} =\phi_l(x_{i,0}, y_{i,0}, \theta_{l_0}, t_0)$ and $\psi_{il}(t=t_0) = \psi_{il,0} =\psi_l(x_{i,0}, y_{i,0}, \theta_{l0}, t_0)$ as the initial potential and stream coordinates of agent $i\in \mathcal{V}_l$ for every $l\in \mathcal{M}$, and
    
    \begin{equation}
        r_{min,0} = \stackunder{min}{\stackunder{i,j}{i$\neq$j}} \sqrt{(\phi_{il,0} - \phi_{jl,0})^2 + (\psi_{il,0} - \psi_{jl,0})^2}
    \end{equation}
    as the minimum separation distance between agents in the $\phi_l-\psi_l$ plane, and
    \begin{equation}
        \zeta_{max} = \max_{x,y} \left(\left({\partial\phi_l\over \partial x}\right)^2 + \left({\partial\phi_l\over \partial y}\right)^2\right).
    \end{equation}
    Assume that the trajectory tracking control error of each individual agent does not exceed $\eta$ and every agent can be enclosed by a ball of radius $\mu$. Then, inter-agent collision avoidance is guaranteed, if the sliding speed $\Dot{\phi}_{il} = v_l$ is the same for every agent in $\mathcal{V}_l$, and
    \begin{equation}
        \label{mainequation}
        \frac{r_{min,0}^2}{\zeta_{max}} \geq 4(\eta+\mu)^2
    \end{equation}
\end{theorem}

\begin{figure}
    \centering
    \includegraphics[width=\linewidth]{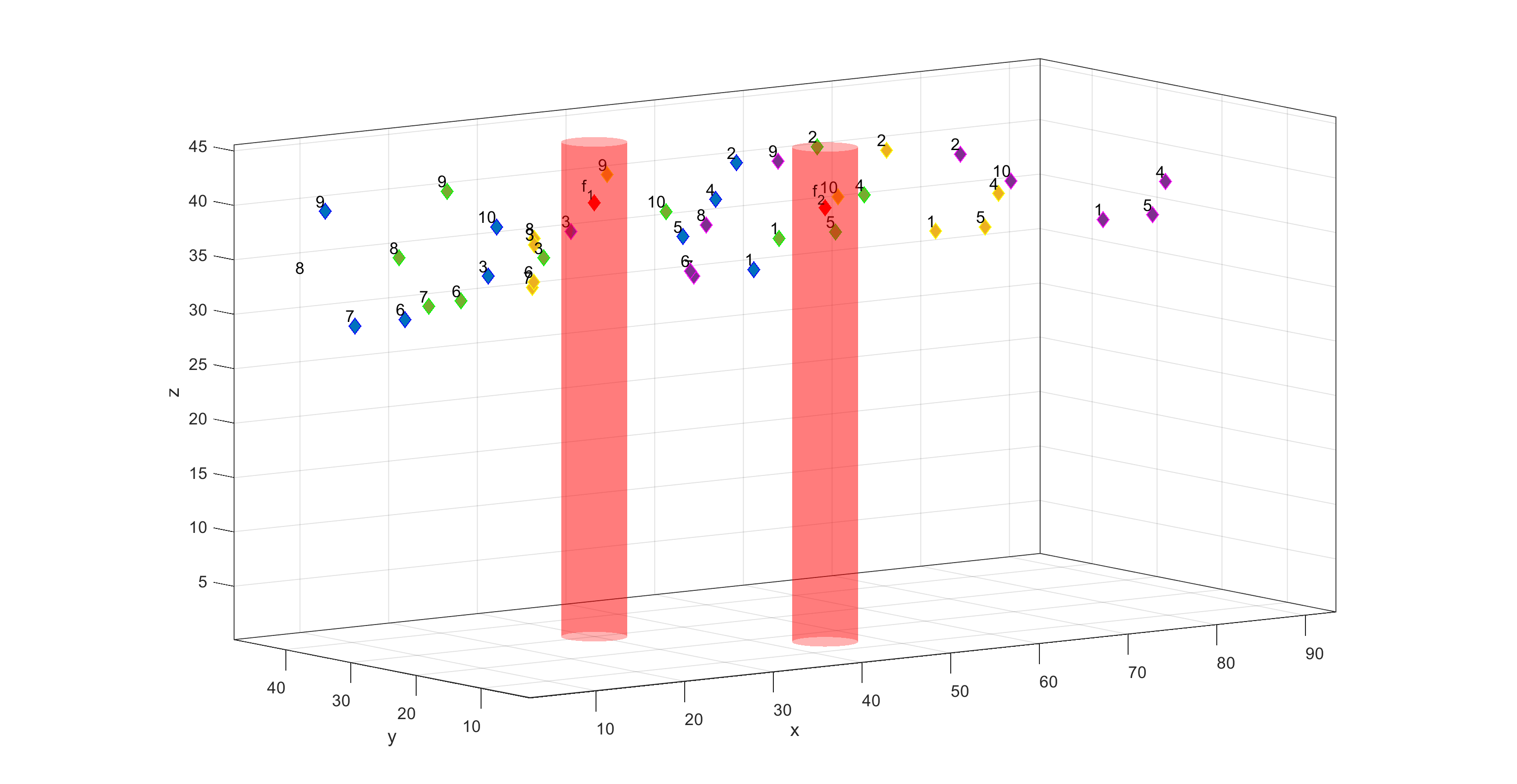}
    \caption{Two-dimensional fluid flow coordination in a $3$-dimensional motion space where obstacles are wrapped by vertical cylinders \cite{emadi2022physics}.}
    \label{fig:acc_2022}
\end{figure}

\begin{proof}
    According to equation (\ref{MainTransformation}), a one-to-one mapping between infinitesimal elements in $(d\phi_l - d\psi_l)$ and $(dx-dy)$ planes exists; they can be related by
    \begin{equation}
    \begin{bmatrix}
        d\phi_l \\
        d\psi_l
    \end{bmatrix}
    = 
    \mathbf{J} \begin{bmatrix}
        dx \\
        dy
    \end{bmatrix}
    \label{eq:1}
\end{equation}
where $\mathbf{J}$ is the Jacobian matrix defined as follows:
    \begin{equation}
        \mathbf{J}(x, y) =
        \begin{bmatrix}
            \frac{\partial \phi_l}{\partial x} & \frac{\partial \phi_l}{\partial y} \\
            \frac{\partial \psi_l}{\partial x} & \frac{\partial \psi_l}{\partial y}
        \end{bmatrix}
    \label{eq:jacobian}
    \end{equation}
     Under Cauchy-Reimann conditions, we can write 
     \[\mathbf{J}^T\mathbf{J} =\left( \left({\partial\phi_l\over \partial x}\right)^2 + \left({\partial\phi_l\over \partial y}\right)^2\right)\mathbf{I}_2,
     \]
     where $\mathbf{I}_2\in \mathbb{R}^{2\times 2}$ is the identity matrix. We can therefore write:
\begin{equation}
    d\phi_l^2 + d\psi_l^2 = \begin{bmatrix}
        dx & dy
    \end{bmatrix} \mathbf{J}^T \mathbf{J} \begin{bmatrix}
        dx \\ 
        dy
    \end{bmatrix}
    =\left(\phi_{lx}^2+\phi_{ly}^2\right)\left(dx^2+dy^2\right).
\end{equation}
If the $x$ and $y$ coordinates along the stream line of every agent $i\in \mathcal{V}_l$ satisfies
    \begin{equation}
        \left(dx^2+dy^2\right) \geq \frac{d\phi_l^2 + d\psi_l^2}{\zeta_{max}}, 
    \end{equation}
then,
    \begin{equation}
        \left(d_{min}(t)\right)^2 \geq \frac{\left(r_{min}(t)\right)^2}{\zeta_{max}}, \qquad \forall t
    \end{equation}
where
    \begin{subequations}
    \begin{equation}
        r_{min}(t) =  \stackunder{min}{\stackunder{i,j}{i$\neq$j}} \sqrt{{\left(\phi_{il}(t) - \phi_{jl}(t)\right)}^2 +\left(\psi_{il}\left(t\right) - \psi_{jl}\left(t\right)\right)^2}, \qquad \forall t,
    \end{equation}
    \begin{equation}\label{dmint}
        d_{min}(t) =  \stackunder{min}{\stackunder{i,j}{i$\neq$j}} \sqrt{{\left(x_{i}(t) - x_{j}(t)\right)}^2 + \left(y_{i}\left(t\right) - y_j\left(t\right)\right)^2}, \qquad \forall t.
    \end{equation}
    \end{subequations}
When the sliding speed $\Dot{\phi_{il}} = v_l$ is the same for every agent $i \in \mathcal{V}_l$, it's agents move as particles of a rigid-body in the $\phi_l-\psi_l$ plane, and thus, inter-agent distances in the $\phi_l-\psi_l$ plane are time-invariant. As a result, the minimum separation distance of the desired formation in the $\phi_l-\psi_l$ plane can be assigned at reference time $t_0$, when the failed agent no-fly zone first appears. Therefore,
$p_{min,0}=p_{min}(t)$ and Eq. \eqref{dminnn} simplifies to 
\begin{equation}\label{dminnn}
    \left( d_{min} (t)\right)^2\geq \frac{\left(r_{min,0}\right)^{2}}{\lambda_{max}},\qquad \forall t
\end{equation}
Since $ d_{min} (t) \geq 2(\eta + \mu)$ is the collision avoidance condition at any time $t$, the inter-agent collision avoidance is assured, if Eq. \eqref{mainequation} is satisfied.
\end{proof}
\begin{remark}
We note that the ideal fluid-flow  coordination is defined over a $2$-D plane, which is called $x-y$ plane in this paper. However,  every agent $i\in\mathcal{V}$ is  free to  move along a direction that is normal to the $x-y$ plane, while $x$ and $y$ components of its desired  position is restricted to slide along a streamline determined by Eq. \eqref{streamline}. For better clarification, Fig. \ref{fig:acc_2022} shows how a multi-agent system, moving in a $3$-D space, can apply the ideal fluid flow model  to safely wrap obstacles specified as vertical cylinders \cite{emadi2022physics}.
\end{remark}

\textbf{Solutions:} Given above problem setting, we will develop \textit{analytic} and \textit{numerical} solutions with the details provided in Sections \ref{Analytic Solution:} and  \ref{Numerical Solution} to define the potential fiction $\phi_l$ and stream function $\psi_l$ for group coordination of $\mathcal{V}_l$'s agents. The analytical approach will be used when $\bar{\mathcal{V}}_l$'s agents are dynamic, and thus, potential and stream curves are time-varying. The analytic method considers $\bar{\mathcal{V}}_l$'s agents as singularity points that are excluded by combining irrational fluid flow patterns.
On the other hand, the numerical solution will be applied to safely plan coordination $\mathcal{V}_l$ in the presence of many static obstacles, with arbitrary size and geometry, that are randomly distributed in the motion space, to maximize the motion space usability while ensuring collision avoidance.

\textbf{ Navigation Modes:} By applying the proposed fluid flow guidance, this paper implements and experimentally evaluates collision-free  navigation of multiple groups of agents under (i) Stationary Non-Concurrent Failures (SNCF), (ii) Time-Varying  Non-Cooperative (TVNC), (iii) Time-Varying Cooperative (TVC), and Stationary Obstacle-Laden Environment (SOLE) scenarios, with the properties listed in Table  \ref{table:Properties}.

\vspace{-0.45cm}
\subsection{Analytic Approach} \label{Analytic Solution:}

 Assuming {\color{black} the agents in }$\mathcal{V}$ {\color{black}operate} in the $x-y$ plane, {\color{black} we represent the }position of agent $i\in \mathcal{V}$ {\color{black} by} complex variable $\mathbf{z}_i=x_i+\mathbf{j}y_i$. {\color{black} In order to} safely plan a collision-free  coordination {\color{black}for agents in} $\mathcal{V}_l$, in the presence of {\color{black} agents in }$\bar{\mathcal{V}}_l$, we treat {\color{black} agents in }${\mathcal{V}}_l$ as a finite number of particles in a time-varying ideal fluid-flow  field{\color{black}. This ideal flow field is defined by combining uniform flow and doublet flow in the $x-y$ plane}. Therefore, {\color{black}agents in} $\mathcal{V}_l$ {\color{black}perceive agents in} $\bar{\mathcal{V}}_l$ {\color{black}as a collection of singularities in} the $x-y$ plane and exclude them by {\color{black}employing} vertical cylinders. {\color{black}These cylinders are derived by} defining {\color{black}the following} complex  function:
\vspace{-0.25cm}
\begin{equation}
    \label{MainTransformation}
    \resizebox{0.99\hsize}{!}{%
$
\begin{split}
\mathbf{f}\left(\mathbf{z}_i\mathrm{e}^{-\mathbf{j}\theta_l(t)},t\right)&=\phi_l(x_i,y_i,\theta_l(t),t)+\mathbf{j}\psi_l(x_i,y_i,\theta_l(t),t)\\
&=\left(1-\beta_l\right)\mathbf{z}_i\mathrm{e}^{-\mathbf{j}\theta_l}+\beta_l\sum_{h\in \bar{\mathcal{V}}_l}\left(\left(\mathbf{z}_i\mathrm{e}^{-\mathbf{j}\theta_l}-{\mathbf{z}}_h(t)\right)+\dfrac{\Delta_h^2}{\mathbf{z}_i\mathrm{e}^{-\mathbf{j}\theta_l}-{\mathbf{z}}_h(t)}\right),
\end{split}
$
}
\end{equation}
{\color{black}This equation applies to} every {\color{black}cluster} $l\in \mathcal{M}$ and {\color{black}agent} $i\in \mathcal{V}_l${\color{black}, where} $\beta_l\in \left\{0,1\right\}$ is a binary variable, $\theta_l(t)$ {\color{black}is a time dependent angle that determines} the bulk motion direction of cluster $l\in \mathcal{M}$, and $\Delta_h\in \mathbb{R}_+$ is {\color{black}the} chosen {\color{black}exclusion radius} such that the size of agent $h\in \bar{\mathcal{V}}_l$ is properly incorporated. {\color{black}Because Eq. \eqref{MainTransformation}} establishes a nonsingular transformation  between $\mathbf{z}_i=x_i+\mathbf{j}y_i$ and $\phi_{il}+\mathbf{j}\psi_{il}$ for every $i\in \mathcal{V}_l$ and $l\in \mathcal{M}$, $(x_i,y_i)$ can be uniquely obtained based on $\left(\phi_{il}(t),\psi_{il}(t)\right)$ at any time $t$ by
\vspace{-0.3cm}
\begin{subequations}
    \begin{equation}\label{xvshs}
        x_i=g_1\left(\phi_{il},\psi_{il},\theta_l,t\right)
    \end{equation}
        \begin{equation}\label{yvshs}
        y_i=g_2\left(\phi_{il},\psi_{il},\theta_l,t\right)
    \end{equation}
\end{subequations}
We use $g_1$ and $g_2$ in Algorithm \ref{alg1} for presenting the position update law.

\begin{table}[]
    \caption{Properties of the investigated fluid-flow navigation problems.}
    \centering
    \begin{tabular}{|c|c|c|c|c|}
     \hline
    Scenarios & $\theta_l~(l\in \mathcal{M})$ & $\mathcal{V}_l~(l\in \mathcal{M})$ & $\beta_l~(l\in \mathcal{M})$&$m$\\
    \hline
    SNCF & Time-Invariant & Time-Varying & $\beta_l=1$ & $m=2$\\
    TVNC & Time-Varying & Time-Invariant & $\beta_l=1$ & $m=2$\\
    TVC & Time-Varying & Time-Invariant & $\beta_l\in \left\{0,1\right\}$ & $m>1$\\
    SOLE & Time-Invariant & Time-Invariant & $\beta_l=1$ & $m=1$\\
    \hline
    \end{tabular}
    \label{table:Properties}
\end{table}

\begin{algorithm}
  \caption{Position Update Algorithm for Every Cluster $l\in \mathcal{M}$ under Fluid-Flow Navigation Strategy.}\label{alg1}
  \begin{algorithmic}[1]
          \State \textit{Get:} Time increment $\Delta t$, $\Delta_h$ and $\mathbf{z}_h$ for every $h\in \bar{\mathcal{V}}_l$, $\beta_l\in \left\{0,1\right\}$, $\theta_l$ sliding speed $v_l$, and current position $\mathbf{z}_i$ of every agent $i\in \mathcal{V}_l$.
         \State \textit{Obtain:} Next position  $\mathbf{z}'_i=x'_i+\mathbf{j}y'_i$.
          \For{\texttt{ $i\in \mathcal{V}_l$}}
           \State Compute current $\phi_{il}\left(t\right)$ and $\psi_{il}(t)$ using Eq. \eqref{MainTransformation}.
           \State Compute next potential  $\phi'_{il}$: $\phi_{il}=\phi_{il}+v_l\Delta t$.
           \State Compute next stream $\psi'_{il}$: $\psi_{il}=\psi_{il}$.
           \State Compute next $x'_i$: $ x'_i=g_1\left(\phi'_{il},\psi'_{il},\theta_l,t\right)$.
           \State Compute next $y'_i$: $ y'_i=g_2\left(\phi'_{il},\psi'_{il},\theta_l,t\right)$.
        \EndFor    
  \end{algorithmic}
\end{algorithm}

\subsection{Numerical Approach}  \label{Numerical Solution}
For the numerical solution, we propose to establish a non-singular mapping between an obstacle-laden ``motion space'', specified by position components $X=x\cos\theta_l+y\sin\theta_l$ and $Y=y\cos\theta_l-x\sin \theta_l$,  and  an obstacle-free ``planning space'' that is defined by coordinates $\phi_l$ and $\psi_l$, where $X+\mathbf{j}Y=\mathbf{z}e^{-\mathbf{j}\theta_l}$ and $\theta_l$ is constant. By using the method presented  in \cite{hoffmann1993computational}, $X(\phi_l,\psi_l )$ and  $Y(\phi_l,\psi_l )$, defined over the planning space, are obtained by solving
\vspace{-0.25cm}
\begin{subequations}\label{eq2}
    \begin{equation}
    a X_{\phi_l \phi_l}-2 b X_{\phi_l \psi_l}+c X_{\psi_l \psi_l}  =0,\qquad l\in \mathcal{M}, 
    \end{equation}
        \begin{equation}
   a Y_{\phi_l \phi_l}-2 b Y_{\phi_l \psi_l}+c Y_{\psi_l \psi_l}  =0,\qquad l\in \mathcal{M}, 
    \end{equation}
\end{subequations}
where $a =X_{\psi_l}^2+Y_{\psi_l}^2$, $b =X_{\phi_l} X_{\psi_l}+Y_{\phi_l} Y_{\psi_l}$, and $c=X_{\phi_l}^2+Y_{\phi_l}^2$. 

\begin{figure}
    \centering
    \includegraphics[width=0.49\textwidth]{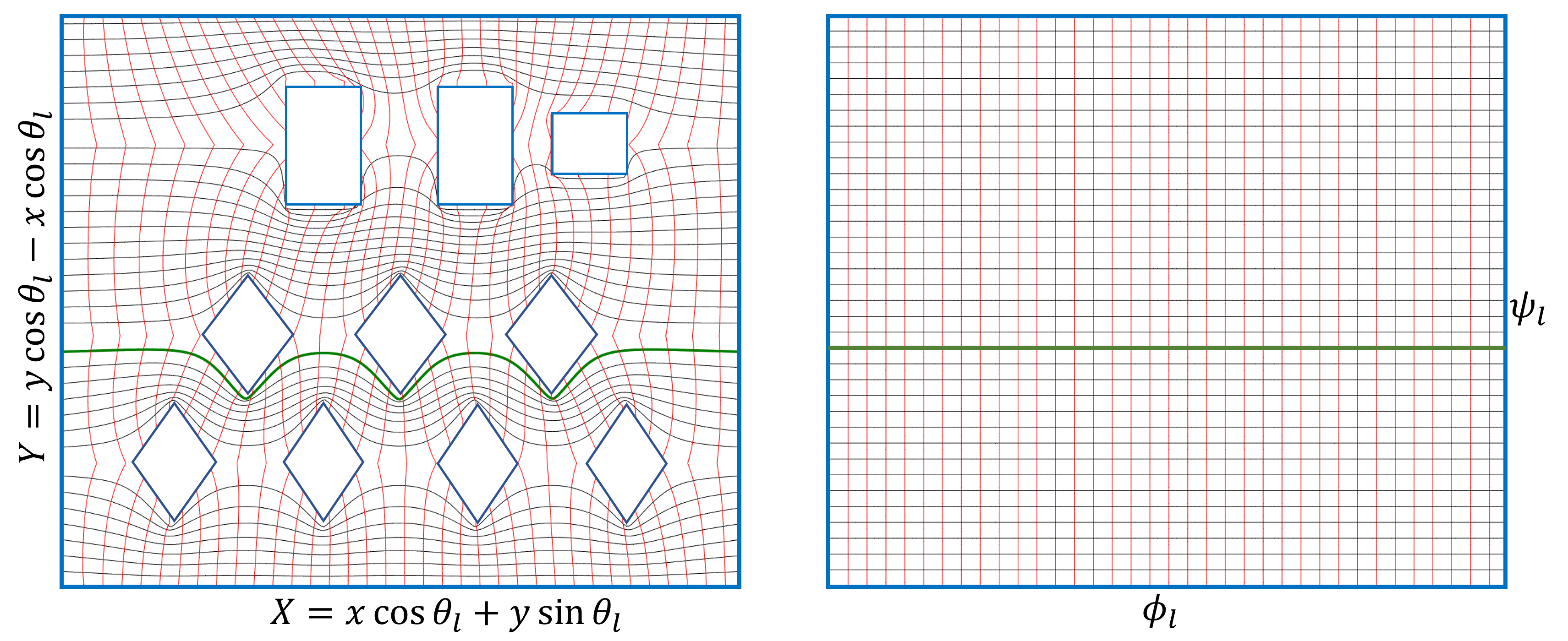}
    \caption{\textit{Left:} Motion space with streamlines shown by black and potential lines shown by red.\textit{ Right:} Planning space. The green curve in the motion space is an streamline used by an agent $i\in \mathcal{V}_l$ to avoid collision with obstacles (the projection of the agent $i$'s path is a horizontal line in the planning space).}
    \label{conversionc}
\end{figure}

For better clarification, Fig. \ref{conversionc}  shows an obstacle-laden environment with eight obstacles. The streamlines shown by the black curves and the potential lines shown by the red curves are both obtained numerically by solving partial differential equations \eqref{eq2} that are defined over the ``planning'' space. As shown in Fig. \ref{conversionc}, an agent following the streamline shown by green can safely avoid an obstacle in the motion space, thus, the path planning strategy presented in Section \ref{sec:methodology} can be used by every agent to safely warp obstacles by sliding along a streamline.

We implement the proposed numerical approach over a rectangular motion space defined by
\vspace{-0.25cm}
\begin{equation}
    \mathcal{P}=\left\{\left(X,Y\right):X\in \left[X_{min},X_{max}\right],Y\in \left[Y_{min},Y_{max}\right]\right\}
\end{equation}
where  $\left(X_{min},Y_{min}\right)$, $\left(X_{min},Y_{max}\right)$, $\left(X_{max},Y_{min}\right)$, and $\left(X_{max},Y_{max}\right)$ are positions of the  $\mathcal{P}$'s corners. Obstacles are defined by subset $\mathcal{O}\subset \mathcal{P}$, and  divided into $n_o$ subsets $\mathcal{O}_1$ through $\mathcal{O}_{n_o}$ ($\mathcal{O}=\mathcal{O}_1\bigcup \cdots \mathcal{O}_{n_o}$). Obstacle subset $\mathcal{O}_j$ consists of finite number of compact obstacle zones  whose $Y$ components of their center of mass are the same and equal to $\bar{Y}_j$. Given $\mathcal{P}$ and $\mathcal{O}$, $\mathcal{N}=\mathcal{P}\setminus \mathcal{O}$ define the navigable space. The navigable space is divided into $n_o+1$ navigable channels $\mathcal{N}_0$, $\cdots$, $\mathcal{N}_{n_o}$, where 
\begin{equation}\label{nj}
    \mathcal{N}_j=\left[X_{min},X_{max}\right]\times \left[\bar{Y}_{j},\bar{Y}_{j+1}\right)-\mathcal{O},\qquad j=0,\cdots,n_o,
\end{equation}
 $\times$ is Cartesian product symbol, $\bar{Y}_0=Y_{min}$, and $\bar{Y}_{n_o+1}=Y_{max}$. 
 
\begin{figure}[ht]
    \centering
    \includegraphics[width=\linewidth]{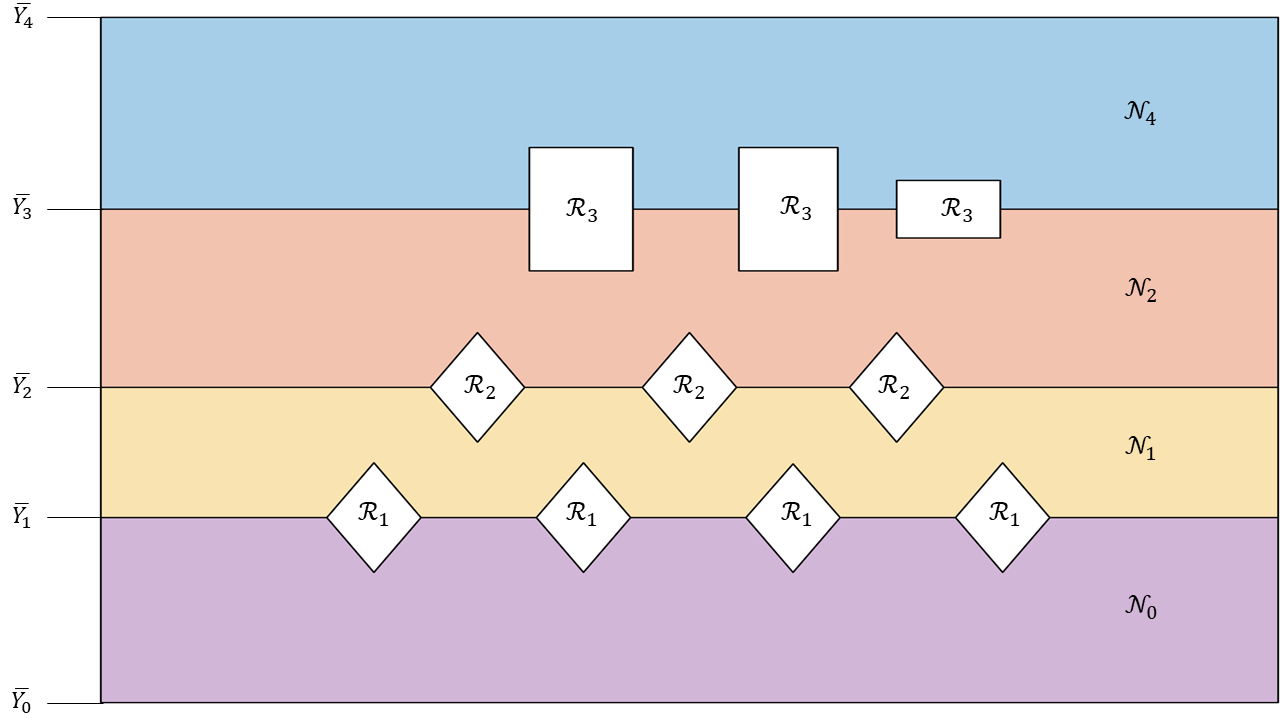}
    \vspace{-0.5cm}
    \caption{The SMART lab floor is defined as the motion space $\mathcal{P}$ and divided into four navigable channels using Eq. \eqref{nj}.}
    \label{fig:game_theory_four_cf_alternate-ffff}
\end{figure}

 For better clarification, we consider the available floor area of the SMART lab   as a rectangular motion space with $X_{min}=Y_{min}=-2.5$ and $X_{max}=Y_{max}=2.5$ (See Fig. \ref{fig:game_theory_four_cf_alternate-ffff}). For the flight experiments, we  consider $8$ cylinders, based by rectangles and diamonds, as static obstacles as shown in Fig. \ref{fig:game_theory_four_cf_alternate}. The obstacles are divided into three group $\mathcal{O}_1$ with $\bar{Y}_1=-1.15$, $\mathcal{O}_2$ with $\bar{Y}_2=-0.20$, and $\mathcal{O}_3$ with $\bar{Y}_3=1.10$. The four navigable channels $\mathcal{N}_0$, $\mathcal{N}_1$, $\mathcal{N}_2$, and $\mathcal{N}_{3}$, obtained Eq. \eqref{nj},  are colored in purple, yellow, red, and blue, respectively. 

\textbf{Solution:} To obtain $\phi_l$ and $\psi_l$ values over $\mathcal{N}$, we first define the boundary of navigable channel $\mathcal{N}_j$, that is denoted by $\partial \mathcal{N}_j$, as follows:
\vspace{-0.25cm}
\begin{equation}
    \partial \mathcal{N}_j=\partial \mathcal{N}_{1,j}\bigcup \partial\mathcal{N}_{2,j}\bigcup\partial \mathcal{N}_{3,j}\bigcup\partial \mathcal{N}_{4,j},\qquad j=1,\cdots,n_o,
\end{equation}
where $\partial \mathcal{N}_{1,j}$, $\partial \mathcal{N}_{2,j}$, $\partial \mathcal{N}_{3,j}$, and $\partial \mathcal{N}_{4,j}$ define the bottom, right, top, and left boundaries of $\mathcal{N}_j$, respectively. 
We uniformly distribute $n_j$ nodes along boundaries $\partial \mathcal{N}_{2,j}$ and $\partial \mathcal{N}_{4,j}$ while $p$ nodes distributed over the boundaries $\partial \mathcal{N}_{1,j}$ and $\partial \mathcal{N}_{3,j}$ are the same for every navigable channel (for every $j\in \left\{0,\cdots,n_{o}\right\}$). As a result, the planning space is also divided into $n_o+1$ rectangles denoted by $\mathcal{S}_0$ through $\mathcal{S}_{n_o}$  where $\partial \mathcal{S}_j$ denotes the boundary of the $j$-th rectangle in the planning space.

\begin{figure}[ht]
    \centering
    \includegraphics[width=\linewidth]{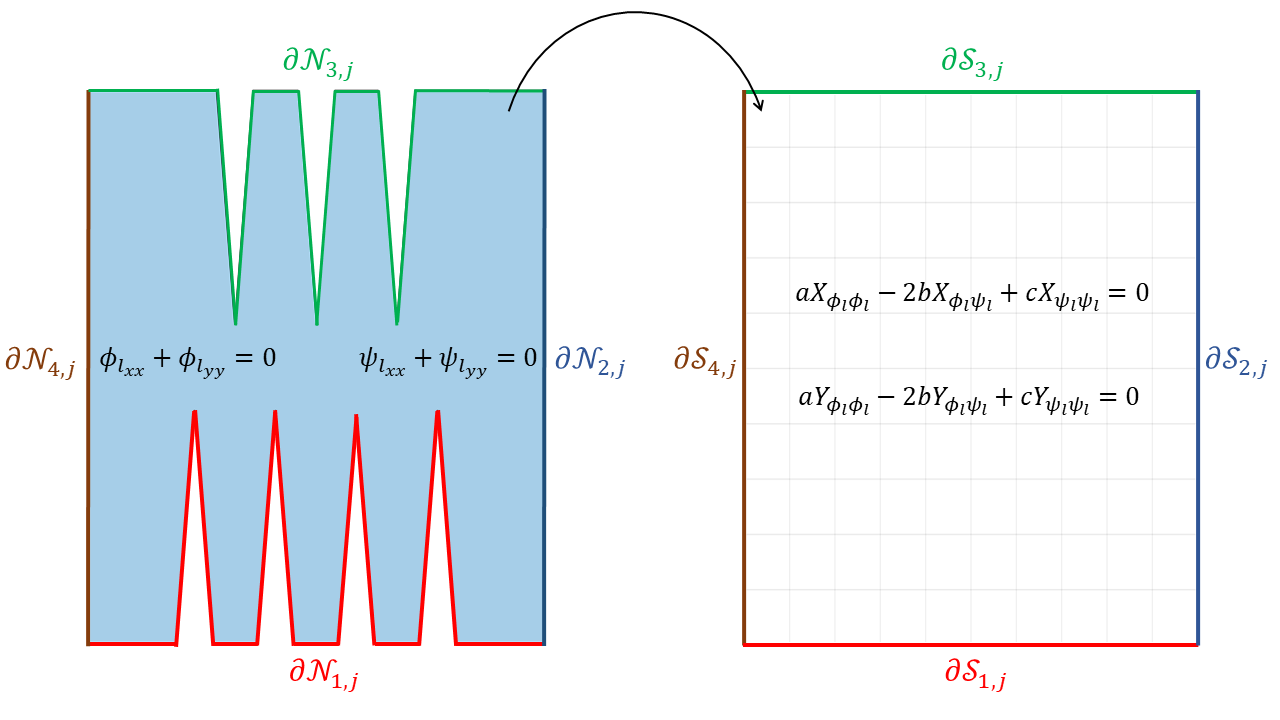}
    \vspace{-0.5cm}
    \caption{Transformation between $\mathcal{N}_j$ (the $j$-th navigable channel) and the $\mathcal{S}_j$. }
    \label{fig:game_theory_four_cf_alternate__1}
\end{figure}

We generate  a uniform grid of size $p\times n_j$ over $\mathcal{S}_j\bigcup \partial \mathcal{S}_j$, for every $j\in \left\{0,\cdots,n_o\right\}$, as shown in Fig. \ref{fig:game_theory_four_cf_alternate__1}.  The boundary conditions of the planning space are then defined as follows:
\vspace{-0.3cm}
\begin{subequations}\label{BC}
\begin{equation}
    X\left(\phi_l,\psi_l\right)=
    \begin{cases}
    \phi_l&\left(\phi_l,\psi_l\right)\in \partial \mathcal{N}_{1,j}\bigcup \partial \mathcal{N}_{3,j}\\
    \phi_{min}&\left(\phi_l,\psi_l\right)\in \partial \mathcal{N}_{4,j}\\
    \phi_{max}&\left(\phi_l,\psi_l\right)\in \partial \mathcal{N}_{2,j}\\
    \end{cases}
\end{equation}
\begin{equation}
    Y\left(\phi_l,\psi_l\right)=
    \begin{cases}
    \psi_l&\left(\phi_l,\psi_l\right)\in \partial \mathcal{N}_{2,j}\bigcup \partial \mathcal{N}_{4,j}\\
    \psi_{min}&\left(\phi_l,\psi_l\right)\in \partial \mathcal{N}_{1,j}\\
    \psi_{max}&\left(\phi_l,\psi_l\right)\in \mathcal{N}_{3,j}\\
    \end{cases}
\end{equation}
\end{subequations}
for  $j=0,\cdots,n_o$, where $\bar{Y}_j\leq \psi_l\leq \bar{Y}_{j+1}$, $\phi_{min}=X_{min}$, $\phi_{max}=X_{max}$, $\psi_{min}=Y_{min}$, and $\psi_{max}=X_{max}$.



\section{ Operation Modes}
\label{Operation Modes}

We use the foundations provided in Section \ref{sec:methodology} to develop algorithms for implementations SNCF, TVNC, TVC, and SOLE operation modes in this section.

\subsection{SNCF  Navigation}
\label{subsec:SNCF}
For the SNCF  navigation model, set $\mathcal{V}$ is divided  into $\mathcal{V}_1$ and $\mathcal{V}_2=\bar{\mathcal{V}}_1$, where $\mathcal{V}_1(t)$ and $\mathcal{V}_2(t)$ are disjoint subsets of $\mathcal{V}$ defining the ``healthy'' and ``faulty'' agents, respectively, at time $t$.

\begin{definition}
    We define $t_{\mathrm{fail}}\geq t_0$ as the \textbf{most recent time} when the status of failure of the agents has changed. 
\end{definition}

For the SNCF coordination, we make the following assumptions:
\begin{assumption}\label{assum1}
 Angle $\theta_1(t)$ is constant at any time $t\geq t_0$, where $t_0$ is the start time of the SNCF coordination.
\end{assumption}
\begin{assumption}
    \label{assum2}
    We assume that either the sliding speed $v_l$, used in Eq. \eqref{Eqqqqqq4}, is sufficiently large, or the geometry of the domain enclosing $h\in \bar{\mathcal{V}}_l$ is spacious enough, such that
    every faulty agent $h\in \mathcal{V}_2$ remains inside a stationary vertical cylinder after its failure is detected.
\end{assumption}

For the SNCF, we define potential and stream functions only for the healthy agents. Thus, potential function $\phi_1$, stream function  $\psi_1$, $\beta_1$, and $\theta_1$ are denoted by $\phi$, $\psi$, $\beta$, and $\theta$, respectively, and substituted in Eq. \eqref{MainTransformation} to compute the potential and stream functions under  SNCF. By imposing Assumptions \ref{assum1} and \ref{assum2}, potential function $\phi(x,y,t_{\mathrm{fail}})$ and stream $\psi(x,y,t_{\mathrm{fail}})$ are piece-wise time-invariant and remains spatially-varying at any time $t\geq t_{\mathrm{fail}}$ until the status of agents' failures change. We apply Algorithm \ref{alg2} to safely plan coordination of healthy agents under the SNCF strategy.

\begin{algorithm}
  \caption{Algorithm for SNCF Fluid-Flow Navigation.}\label{alg2}
  \begin{algorithmic}[1]
        \State \textit{Get:} Initial time $t_0$, number of time steps denoted by $n$, time increment $\Delta t$,  healthy agent set $\mathcal{V}_1(t_0)$, faulty agent set ${\mathcal{V}}_2(t_0)$,
          time increment $\Delta t$, $\Delta_h=\Delta$ for every $h\in {\mathcal{V}}_2$, $\theta_l$, $v_l$, initial position $\mathbf{z}_i(t_0)$ of every healthy agent $i\in \mathcal{V}_1$, and initial position $\mathbf{z}_h(t_0)$ of every faulty agent $h\in \mathcal{V}_2$
          \State \textit{Set:} $\beta=1$, $k=1$, $t_h=t_0$.
           \While{$k < n$}
           \If{$\mathcal{V}_2\left(t_k\right)\neq \mathcal{V}_2\left(t_{k-1}\right)$}           
               \State $t_{\mathrm{fail}}\leftarrow t_k$
               \State Update  $\phi\left(x,y,t_{\mathrm{fail}}\right)$.
               \State Update  $\psi\left(x,y,t_{\mathrm{fail}}\right)$.
           \EndIf 
           \State $\phi\left(x,y,t_k\right)\leftarrow \phi\left(x,y,t_{\mathrm{fail}}\right)$.
            \State $\psi\left(x,y,t_k\right)\leftarrow \psi\left(x,y,t_{\mathrm{fail}}\right)$.
           \State Obtain $\mathbf{z}_i(t_{k+1})$ for every $i\in \mathcal{V}_1(t_k)$ by Algorithm \ref{alg1}.
           \State Return $\mathbf{z}_i(t_{k+1})$.
           \State $\mathbf{z}_i(t_{k})\leftarrow \mathbf{z}_i(t_{k+1})$ for every $i\in \mathcal{V}_1(t_k)$.     
           \State $k\leftarrow k+1$.        
       \EndWhile
  \end{algorithmic}
\end{algorithm}

\subsection{TVNC  Navigation}
\label{subsec:TVNC}
For the TVNC  navigation model, set $\mathcal{V}$ is divided  into time-invariant subsets $\mathcal{V}_1$ and $\mathcal{V}_2=\bar{\mathcal{V}}_1$, where $\mathcal{V}_1$ and $\mathcal{V}_2$ define ``cooperative'' and ``non-cooperative'' agents, respectively. The noncooperative agents have a predefined trajectories in the motion space whereas the cooperative agents uses the  navigation model, presented in Algorithm \ref{alg1}, to safely update their positions and reach their target positions.

Similar the SNCF coordination model, we denote potential function $\phi_1$, stream function  $\psi_1$, $\beta_1$, and $\theta_1$ by $\phi$, $\psi$, $\beta$, and $\theta$, respectively, and substitute them in Eq. \eqref{MainTransformation} to compute the potential and stream functions. Let $\mathbf{z}_{i,f}=x_{i,f}+\mathbf{j}y_{i,f}$ be the known target position of cooperative agent $i\in \mathcal{V}_1$, then, then angle $\theta$, assigning the bulk motion direction of $\mathcal{V}_1$ is obtained by
\vspace{-0.25cm}
\begin{equation}\label{thetasecond}
    \theta(t)=\theta_l=\arg \left(\sum_{i\in \mathcal{V}_1}\left(\mathbf{z}_{i,f}-\mathbf{z}_i(t)\right)\right),\qquad \forall t\geq t_0,~l\in \mathcal{M},
\end{equation}
where $t_0$ is the initial time. We use Algorithm \ref{alg3} to safely plan coordination of every agent  $i\in \mathcal{V}$ in a shared motion space.

\begin{algorithm}
  \caption{Algorithm for TVNC Fluid-Flow Navigation.}\label{alg3}
  \begin{algorithmic}[1]
        \State \textit{Get:} Initial time $t_0$, time increment $\Delta t$, $\epsilon>0$,  $\Delta_h$ for every $h\in \bar{\mathcal{V}}_l$, $\theta_l$, $v_l$, initial position $\mathbf{z}_i(t_0)$ of every cooperative agent $i\in \mathcal{V}_1$.
          \State \textit{Set:} $\beta=1$, $k=0$.
           \While{$\sum_{i\in \mathcal{V}_1}\left|\mathbf{z}_i(t_k)-\mathbf{z}_{i,f}\right|> \left|\mathcal{V}_1\right|\epsilon$ for every $i\in \mathcal{V}_1$}
           \State Get $\mathbf{z}_h\left(t_k\right)$ for every $h\in \mathcal{V}_2$.
           \State Compute $\theta(t_k)$ by Eq. \eqref{thetasecond}.
           \State Update $\phi\left(x,y,\theta(t_k),t_k\right)$ and $\psi\left(x,y,\theta(t_k),t_k\right)$.
           \State Obtain $\mathbf{z}_i(t_{k+1})$ for every $i\in \mathcal{V}_1$ by Algorithm \ref{alg1}.
           \State Return $\mathbf{z}_i(t_{k+1})$.
           \State $\mathbf{z}_i(t_{k})\leftarrow \mathbf{z}_i(t_{k+1})$ for every $i\in \mathcal{V}_1$.     
           \State $k\leftarrow k+1$.        
       \EndWhile
  \end{algorithmic}
\end{algorithm}

\subsection{TVC  Navigation}
\label{subsec:TVC}
For the TVC  navigation, we enable every agent $i\in \mathcal{V}_l$ to check if there is a possibility of colliding with an agent $h\in \bar{\mathcal{V}}_l$ within the next $n_\tau$ time steps. To this end, we define virtual box $\mathcal{B}_i(t)\subset \mathbb{C}$ for every agent $i\in \mathcal{V}_l$, with side lengths $2\delta$ and $v_ln_\tau
\Delta t$, to check possibility of collision with an agent $j\in \bar{\mathcal{V}}_l$ within the next $n_\tau \Delta t$ seconds. To formally specify collision avoidance condition, we define condition $\zeta$ as follows:
\begin{equation}\label{motionspacecontainment}
\displaystyle\bigvee_{l\in \mathcal{M}}\displaystyle\bigvee_{i\in \mathcal{V}_l}\bigvee_{j\in \bar{\mathcal{V}}_l}\left(\mathbf{z}_j\in \mathcal{B}_i\right),
\tag{$\zeta$}
\end{equation}
where ``$\bigvee$'' is used to specify ``at least one''. 
\begin{figure}
    \centering
    \includegraphics[width=0.4\textwidth, height=5cm]{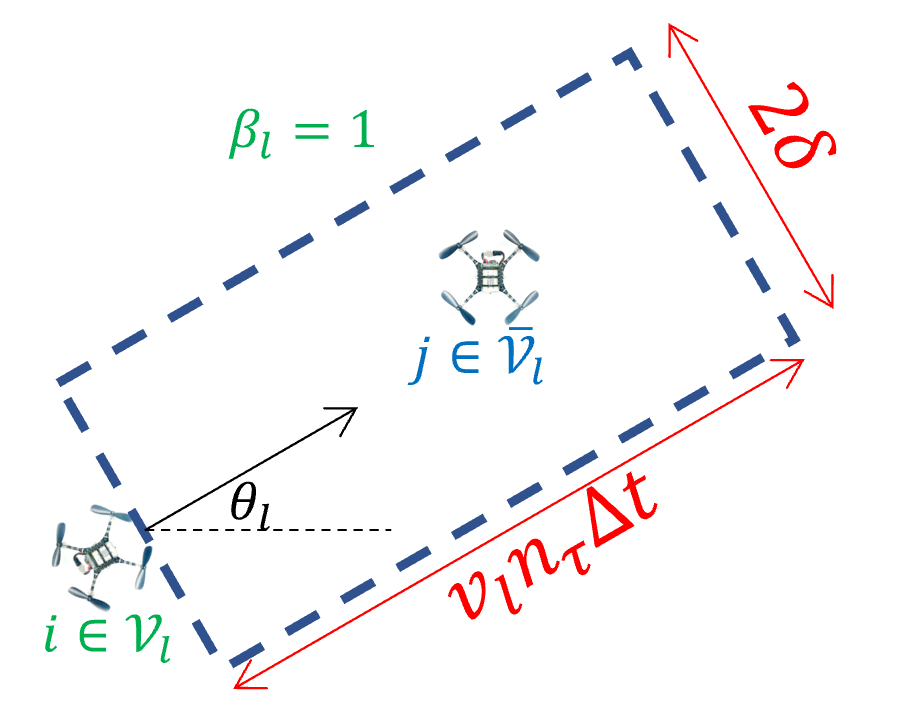}
    \vspace{-0.25cm}
    \caption{The virtual box $\mathcal{B}_i$ with side lengths $2\delta$ and $v_ln_\tau\Delta t$ used by agent $i\in \mathcal{V}_l$ to estimate the possibility of colliding an agent $j\in \bar{\mathcal{V}}_l$.}
    \label{connectionschematic}
\end{figure}
Note that $\zeta$ is satisfied, if there exists at least one agent $j\in \bar{\mathcal{V}}_l$ that is inside one of the safety boxes of $\mathcal{V}_l$'s agents. 

Therefore, $\beta_l$ is specified as follows:
\begin{subequations}
    \begin{equation}
        \zeta \implies \bigwedge_{l\in \mathcal{M}}\left(\beta_l=1\right),
    \end{equation}
        \begin{equation}
        \neg \zeta \implies \bigwedge_{l\in \mathcal{M}}\left(\beta_l=0\right),
    \end{equation}
\end{subequations}
where  ``$\bigwedge$'' is used to specify ``include all''; ``$\implies$'' means ``implies that''; and ``$\neg$''  is the ``negation'' symbol. 
For the TVC, we use Algorithm 4 to safely plan coordination of every agent  $i\in \mathcal{V}$ in a shared motion space.

\begin{algorithm}
  \caption{Algorithm for TVC Fluid-Flow Navigation.}
  \label{alg4}
  \begin{algorithmic}[1]
        \State \textit{Get:} Initial time $t_0$, number of sample times denoted by $n$, time increment $\Delta t$,  $\Delta_h$ for every $h\in \bar{\mathcal{V}}_l$, $v_l$, $\delta$, $n_\tau$, initial position $\mathbf{z}_i(t_0)$ of every  agent $i\in \mathcal{V}_l$ and every cluster $l\in\mathcal{M}$.
          \State \textit{Set:}  $k=0$.
          \State \textit{Set:}  $\beta_l(t_0)=0$, $\cdots$, $\beta_l(t_n)=0$ for every $l\in \mathcal{M}$.
          \For{\texttt{ $k\in \left\{0,\cdots,n\right\}$}}
          \For{\texttt{ $l\in \mathcal{M}$}}
          \If{$\zeta$~ is satisfied}           
               \State $\beta_l\left(t_k\right)=1$.
               \State Get $\mathbf{z}_h\left(t_k\right)$ for every $h\in \bar{\mathcal{V}}_l$.
           \EndIf 
           \State  Compute $\theta_l(t_k)$ by Eq. \eqref{thetasecond}.            
           \State Update $\phi_l\left(x,y,\theta(t_k),t_k\right)$ and $\psi_l\left(x,y,\theta(t_k),t_k\right)$.
           \State Obtain $\mathbf{z}_i(t_{k+1})$ for every $i\in \mathcal{V}_l$ by Algorithm \ref{alg1}.
           \State Return $\mathbf{z}_i(t_{k+1})$.
           \State $\mathbf{z}_i(t_{k})\leftarrow \mathbf{z}_i(t_{k+1})$ for every $i\in \mathcal{V}_l$.                 
        \EndFor 
        \State $k\leftarrow k+1$.      
        \EndFor  
  \end{algorithmic}
\end{algorithm}

\subsection{SOLE  Navigation}
\label{subsec:SOLE}
For the SOLE navigation, we consider operation of a single agent team in an obstacle-laden environment, thus, $m=1$ and $\mathcal{V}=\mathcal{V}_1=\left\{1,\cdots,N\right\}$ defines the identification numbers of the agents, $\phi(x,y)=\phi_1(x,y)$ and $\psi(x,y)=\psi_1(x,y)$ denote the potential and stream function, and $\theta=\theta_1$ is constant.  We propose the Algorithm \ref{alg5} to obtain safe trajectories of every $i\in \mathcal{V}$ by following the motion strategy presented in Section \ref{sec:methodology}. Note that ``C$\theta$'' and ``S$\theta$'' in line $15$ of Algorithm \ref{alg5} stand for ``$\cos\theta$'' and ``$\sin\theta$'', respectively. 

\begin{algorithm}
  \caption{Algorithm for SOLE Fluid-Flow Navigation}
  \label{alg5}
  \begin{algorithmic}[1]
        \State \textit{Get:} Initial time $t_0$, number of sample times denoted by $n$, time increment $\Delta t$, $\mathcal{O}_1$ through $\mathcal{O}_{n_o}$, initial positions $\mathbf{z}_{i,0}=x_{i,0}+\mathbf{j}y_{i,0}$ of every $i\in \mathcal{V}$, $\phi_{min}$, $\phi_{max}$, $p$, and $\theta$.
          \State \textit{Set:}  $k=0$.
          \State $\Delta\phi=\left(\phi_{max}-\phi_{min}\right)/p$.
          \State Determine navigable channels by Eq. \eqref{nj}.
          \State Specify boundary conditions using Eq. \eqref{BC}.
          \State Obtain $X\left(\phi,\psi\right)$ and $Y(\phi,\psi)$ over $\mathcal{P}$ numerically,  by solving Eq. \eqref{eq2}.
          \State Obtain $X_{i,0}=x_{i,0}\cos \theta+x_{i,0}\sin \theta$ for every $i\in \mathcal{V}$.
          \State Obtain $Y_{i,0}=y_{i,0}\cos \theta-x_{i,0}\sin \theta$ for every $i\in \mathcal{V}$.
          \State Compute associated $\left(\phi_{i}(t_0),\psi_{i}(t_0)\right)\in \mathcal{S}$ for every $i\in \mathcal{V}$.
          \For{\texttt{ $k\in \left\{0,\cdots,n\right\}$}}
            \For{\texttt{ $i\in \mathcal{V}$}}
                \State $\phi_i\left(t_{k+1}\right)\leftarrow \phi_i\left(t_{k}\right)+\Delta\phi$.
                \State $\psi_i\left(t_{k+1}\right)\leftarrow \psi\left(t_{k}\right)$.
                \State Obtain $(X_i,Y_i)$ associated with  $(\phi_i,\psi_i)$.
                \State $\mathbf{z}_i\left(t_{k+1}\right)\leftarrow(X_iC \theta-Y_i S \theta)+\mathbf{j}(X_iS \theta+Y_i C \theta)$.
                 \State Return $\mathbf{z}_i(t_{k+1})$.
                \State $\mathbf{z}_i(t_{k})\leftarrow \mathbf{z}_i(t_{k+1})$ for every $i\in \mathcal{V}$.
            \EndFor 
        \State $k\leftarrow k+1$.      
        \EndFor  
  \end{algorithmic}
\end{algorithm}

\section{Experiments and Discussion}
\label{sec:ExperimentsAndDiscussion}

We experimentally evaluate the performance of the proposed algorithms and validate the results on a group of tiny quadcopters ({\color{black}The multimedia} of our experiments is available at YouTube {\color{black}(\href{https://youtu.be/4PerPdyj6vM}{Link})}).


\begin{figure}
    \centering
    \includegraphics[width=\linewidth]{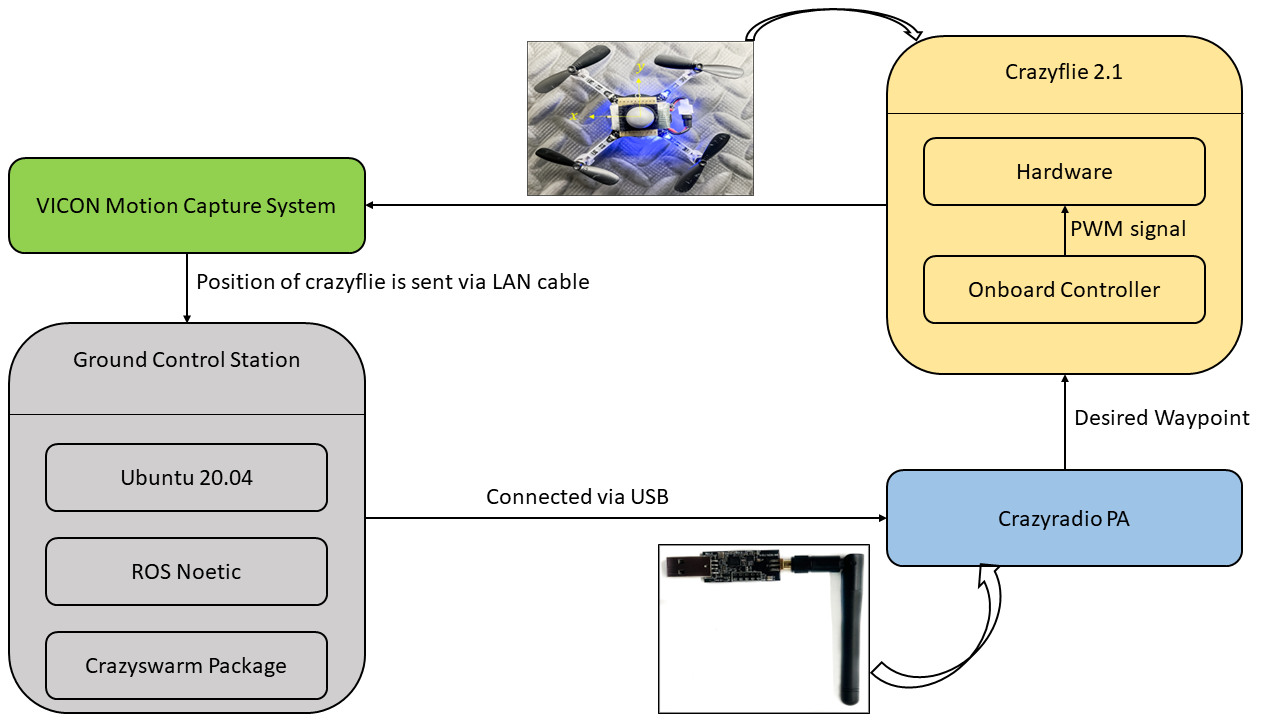}
    \caption{An overview of the experiment setup.}
    \label{fig:experiment_setup}
\end{figure}


The experimental setup
includes $4$ major components shown in Fig. \ref{fig:experiment_setup}:
(i) Motion Capture System (MCS), (ii)  Ground Control Station (GCS), (iii) Crazyradio PA, and (iv) Crazyflie $2.1$. The MCS captures the position of the each crazyflie in $3$-D space and sends the information to the GCS through an ethernet cable at {100}{Hz}. The GCS is an Intel i$7$ $11$-th gen desktop, with $16$ GB of RAM running Ubuntu $20.04$ and ROS Noetic. GCS is also installed with the Crazyswarm \cite{preiss2017crazyswarm} ROS stack built by USC-ACT Lab and acts as a centralized planner for the system. The GCS uses the information from MCS to compute the desired states for each crazyflie and then transmits the data to the onboard controller through Crazyradio PA. 
We conducted flight tests at the University of Arizona's Scalable Move and Resilient Transversability (SMART) lab's indoor flying area with a volume of {5}{m} $\times$ {5}{m} $\times$ {2}{m} equipped with $8$ VICON motion capture cameras.
We assume that all crazyflies are flying at the altitude of \SI{1}{m}.

\subsection{Stationary Non-Concurrent Failures (SNCF) Experiment}

\begin{figure}[ht]
    \centering
    \subfigure[]{\includegraphics[width=0.48\linewidth]{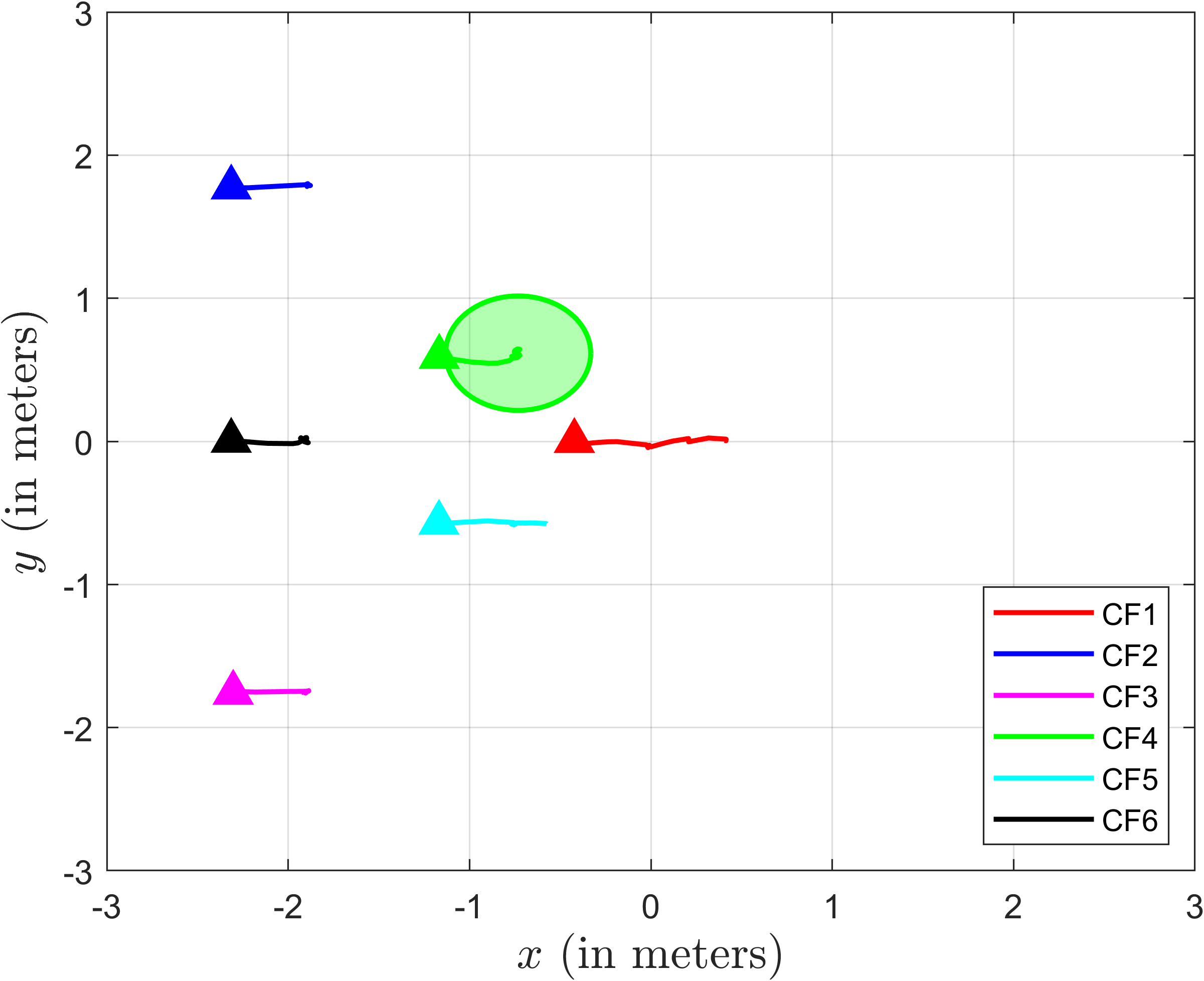}}
    \subfigure[]{\includegraphics[width=0.48\linewidth]{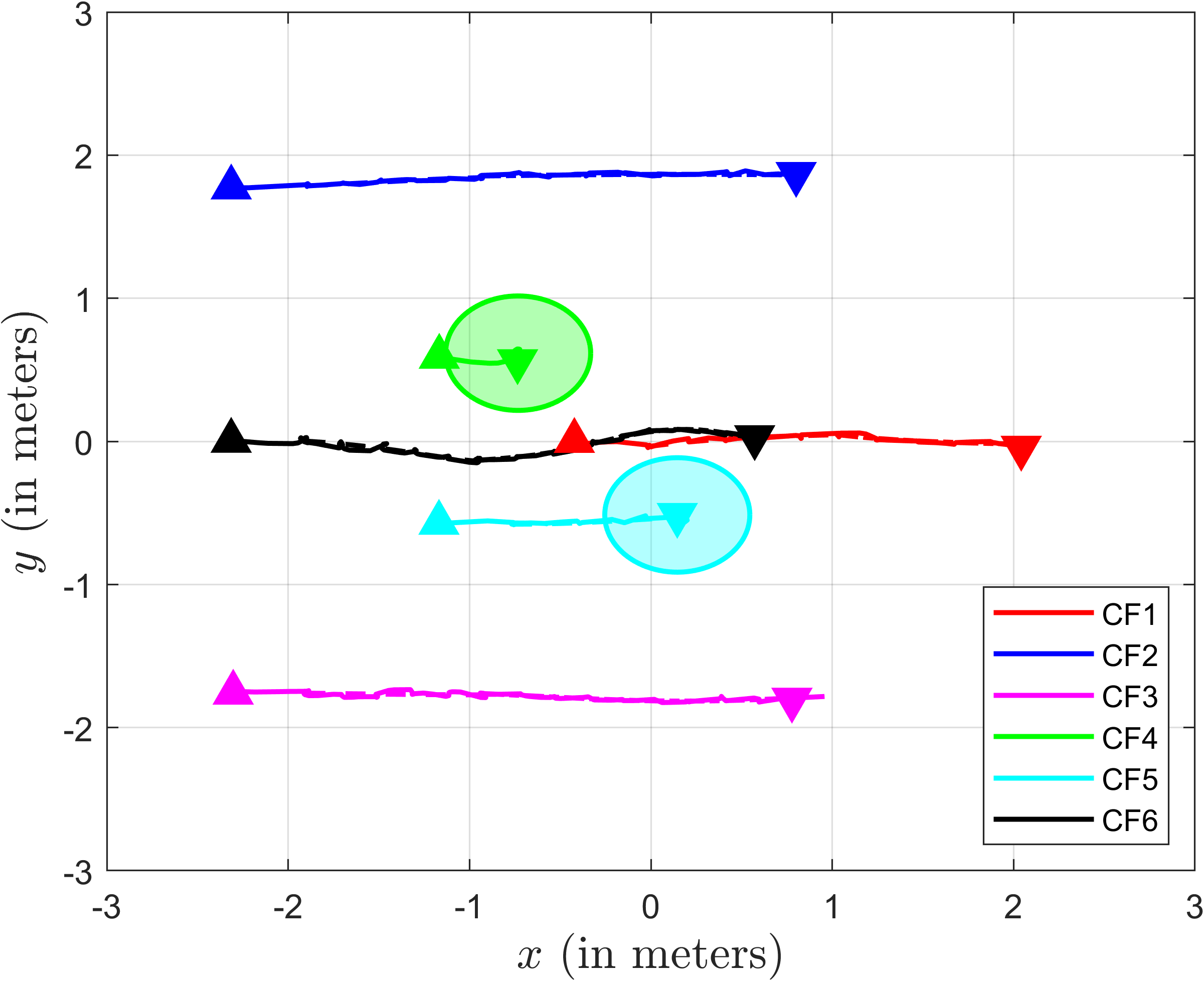}}
    \subfigure[]{\includegraphics[width=0.48\linewidth]{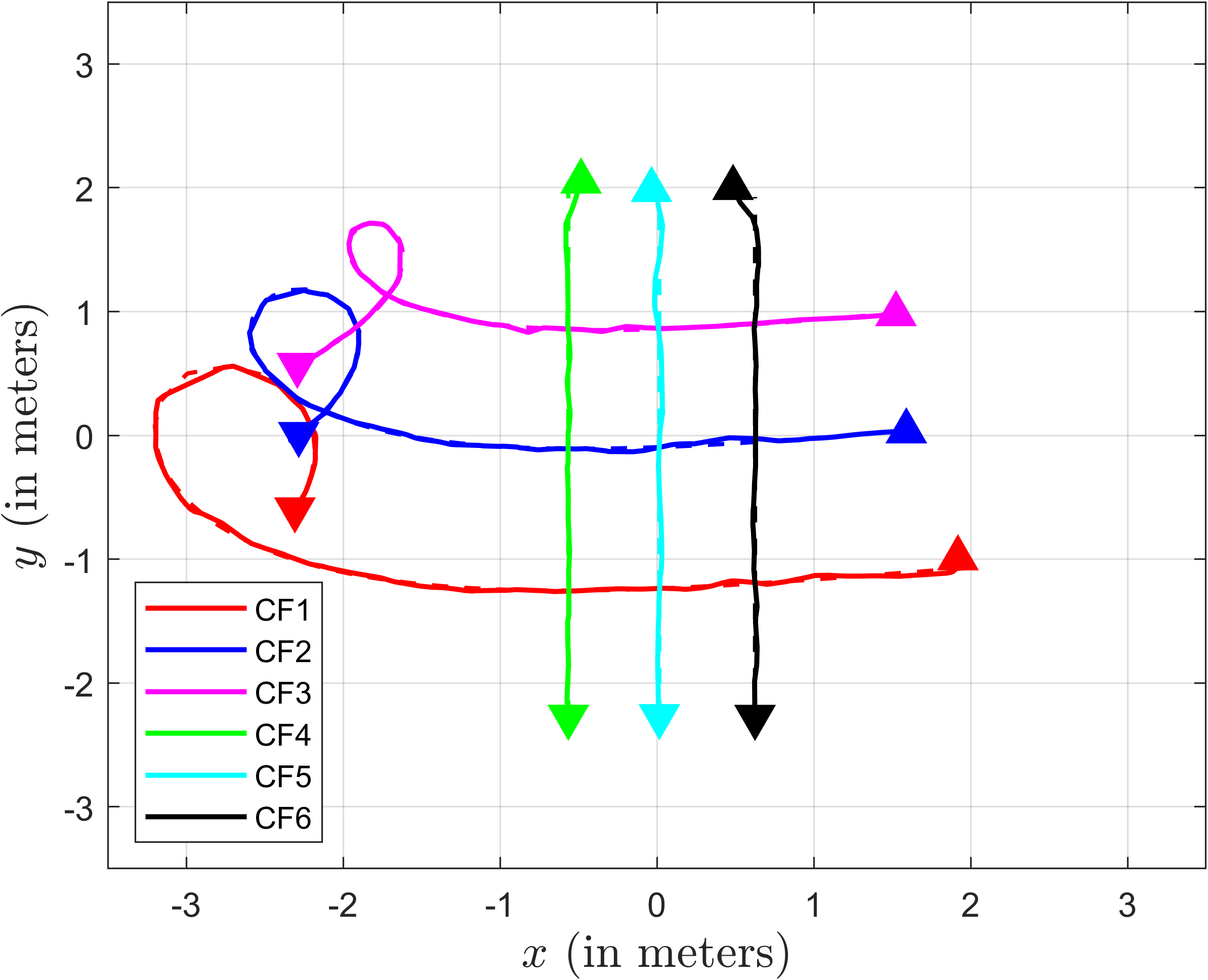}}
    \subfigure[]{\includegraphics[width=0.48\linewidth]{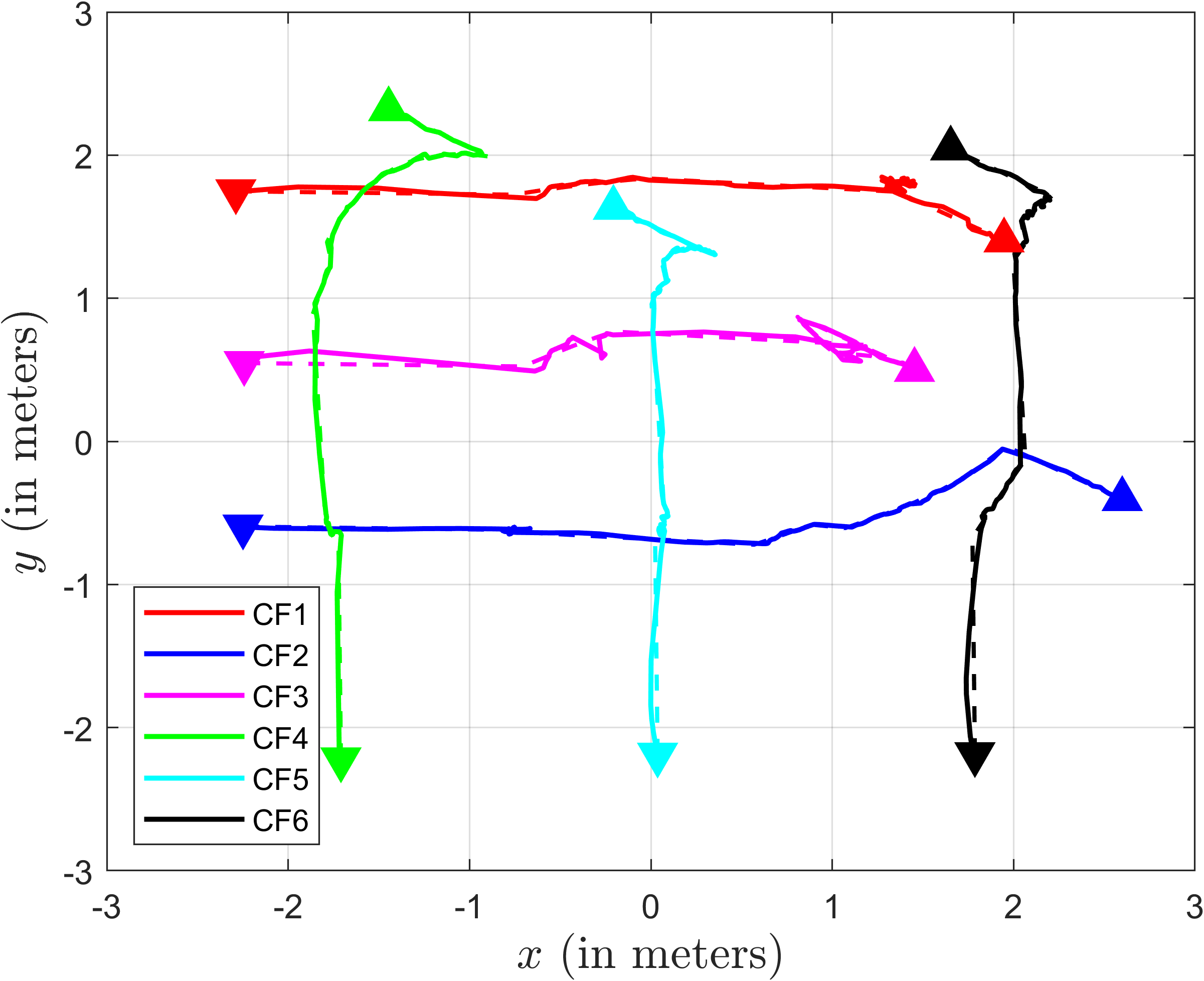}}
    \vspace{-0.3cm}
    \caption{(a) Location of healthy CFs at the time of first failure. The green circle corresponds to the unsafe-zone of CF$4$. (b) Desired (dashed line) versus actual (solid line) paths tracked by CF to avoid unsafe-zones. (c) Desired (dashed line) vs actual (solid line) paths undertaken by CFs. We can see $\mathcal{V}_1$ taking advantage of the  recovery algorithm in order to avoid any collision with non-cooperative CFs in set $\mathcal{V}_2$. (d) Desired (dashed line) vs actual (solid line) paths tracked by CFs using algorithm \ref{alg4} when $N = 6$.}
    \label{fig:sncf_failures}
\end{figure}

For this experiment, we follow the approach presented in Section \ref{subsec:SNCF} where $\Delta_h = $ {0.4}{m} (Eq. \eqref{MainTransformation}). The CFs are uniquely identified using the set $\mathcal{V}=\{1,\cdots,6\}$. As indicated in Table \ref{table:Properties}, we have $m=2$. At time $t_0$, {\color{black} $\mathcal{M} = \{1,2\}$,} $\mathcal{V}$ is divided into $\mathcal{V}_1 = \{1,\cdots,6\}$ and $\mathcal{V}_2 = \emptyset$. Until the first failure, the CFs all move together represented as solid lines (See Fig. \ref{fig:sncf_failures}(a)). At $t_{\mathrm{fail}_1} =$ {2}{s}, CF$4$, chosen randomly, is subjected to failure and is wrapped by a green cylinder representing the unsafe zone (See Fig. \ref{fig:sncf_failures}(a)). At this instant, $\mathcal{V}_1 = \{1,2,3,5,6\}$ and $\mathcal{V}_2 = \mathcal{V} \setminus \mathcal{V}_1 = \{4\}$. The desired paths for all CFs belonging to $ \mathcal{V}_1$ are computed based on algorithm \ref{alg2}. 
We deploy another failure, CF$5$, in the system at $t_{\mathrm{fail}_2} =$ {12}{s} (See Fig. \ref{fig:sncf_failures}(b)). The sets $\mathcal{V}_1$ and $\mathcal{V}_2$ are updated: $\mathcal{V}_1 = \{1,2,3,6\}$ and $\mathcal{V}_2 = \{4,5\}$. The desired paths for the healthy CFs are again computed until $ \mathcal{V}_1$'s CFs have completely passed the unsafe-zones (See Fig. \ref{fig:sncf_failures}(b)).

\vspace{-2mm}
\subsection{Time-Varying Non-Cooperative (TVNC) Experiment}
In this experiment, we evaluate the performance of the algorithm \ref{alg3} proposed in Section \ref{subsec:TVNC} using $m=2$ groups of crazyflies. Specifically, at time $t_0$, the set $\mathcal{V}=\{1,\cdots,6\}$ is divided into time-invariant subsets $\mathcal{V}_1=\{1,2,3\}$ and $\mathcal{V}_2=\{4,5,6\}$. The aim of agents in $\mathcal{V}_2$, known as non-cooperative agents, is to reach their goal locations quickly. Therefore, trajectories of $\mathcal{V}_2$'s agents are predefined, as indicated in Fig. \ref{fig:sncf_failures}(c) (See the green, cyan, and black paths). However, the agents belonging to $\mathcal{V}_1$, termed as cooperative agents, use the fluid-flow navigation function to safely plan paths in the shared motion space. As shown in Fig. \ref{fig:sncf_failures}(c), cooperative CFs $1$, $2$, and $3$ reach their target locations by following the red, blue, and pink paths. 
\begin{figure}[h]
    \centering
    \includegraphics[width=\linewidth]{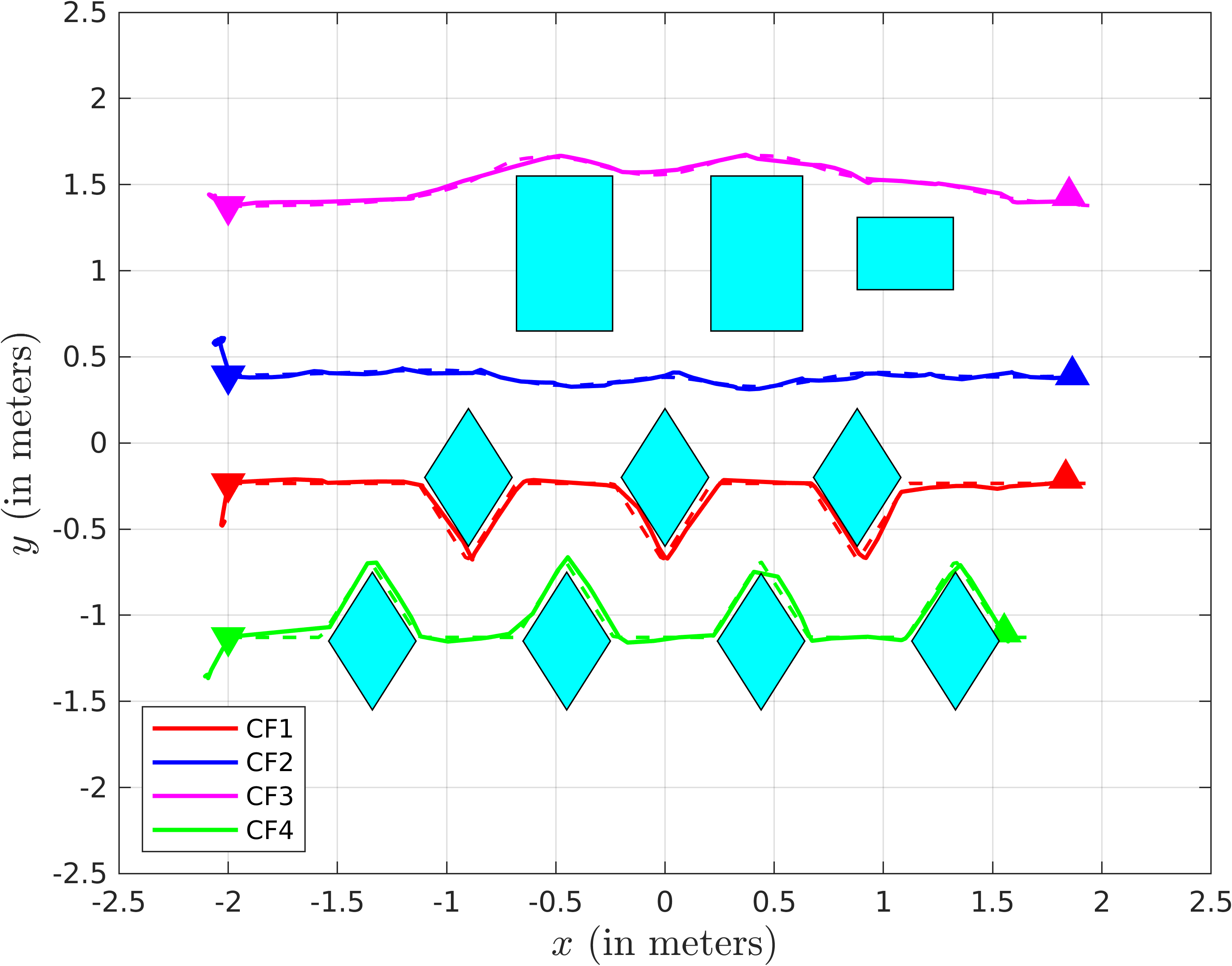}
    \vspace{-0.5cm}
    \caption{Desired (dashed line) vs Actual (solid line) paths tracked by crazyflies using Algorithm  when $N = 4$.}
    \label{fig:game_theory_four_cf_alternate}
\end{figure}

\subsection{Time-Varying Cooperative (TVC)}
According to Section \ref{subsec:TVC} and \ref{alg4}, we define the set $\mathcal{V}=\{1,\cdots,6\}$ to uniquely identify all CFs. CFs are divided into two groups identified by  $\mathcal{V}_1 = \{1,2,3\}$ and $\mathcal{V}_2 = \{4,5,6\}$. For this experiment, we choose $v_l = $ {0.3}{m/s}, $\delta = 0.15$, and $n_{\tau} = 3$. We have experimentally evaluated the scenario when $\theta$ is varying in time but constant for each individual group of agents. This approach ensures that each agent in a group move parallel to other agents in the group. Figure \ref{fig:sncf_failures}(d) plots the result of our experiment. 


\subsection{Stationary Obstacle-Laden Environment (SOLE) Experiment}

In this experiment, we have an obstacle-laden environment as shown in Figure \ref{fig:game_theory_four_cf_alternate}. We follow the approach presented in Section \ref{subsec:SOLE} and define the set $\mathcal{V} = \{1,2,3,4\}$. By implementing Algorithm  \ref{alg5}, CF quadcopters  $1$ through $4$ follow the paths shown in Fig. \ref{fig:game_theory_four_cf_alternate} to safely pass through obstacles.


\section{CONCLUSION}
\label{sec:conclusion}
In this work, we proposed {\color{black}multiple} recovery algorithms based on ideal fluid-flow for collision-free coordination between multiple groups of agents. Our algorithms were able to handle different scenarios including stationary non-concurrent failures, time-varying non-cooperative failures, and time-varying cooperative failures. Experimental results using teams of crazyflies displayed the advantage of our proposed algorithms in handling different situations. Future work on this direction include incorporating reinforcement learning techniques.

\bibliographystyle{IEEEtran}
\bibliography{reference.bib}

\end{document}